\documentclass[final,authoryear,3p,times,twocolumn]{elsarticle}

\usepackage{amssymb}

\journal{Applied Radiation and Isotopes}

\begin{document}

\begin{frontmatter}



\title{Proton-induced cross-sections of nuclear reactions on lead up to 37 MeV}


\author[1]{F. Ditr\'oi\corref{*}}
\author[1]{F. T\'ark\'anyi}
\author[1]{S. Tak\'acs}
\author[2]{A. Hermanne}
\cortext[*]{Corresponding author: ditroi@atomki.hu}

\address[1]{Institute for Nuclear Research, Hungarian Academy of Sciences (ATOMKI),  Debrecen, Hungary}
\address[2]{Cyclotron Laboratory, Vrije Universiteit Brussel (VUB), Brussels, Belgium}

\begin{abstract}
Excitation function of proton induced nuclear reactions on lead for production of  $^{206,205,204,203,202,201g}$Bi, $^{203cum,202m,201cum}$Pb and $^{202cum,201cum,200cum,199cum}$Tl radionuclides  were measured up to 36 MeV by using activation method, stacked foil irradiation technique and ?-ray spectrometry. The new experimental data were compared with the few earlier experimental results and with the predictions of the EMPIRE 3.1, ALICE-IPPE (MENDL2p) and TALYS (TENDL-2012) theoretical reaction codes.
\end{abstract}


\end{frontmatter}


\section{Introduction}
\label{1}
Production cross-sections of proton induced nuclear reactions on lead are important for many applications and for the development of nuclear reaction theory. Lead is an important technological material as pure element as well as alloying agent and it is also widely used in different nuclear technologies, as well as target material for production of $^{201}$Tl medical diagnostic gamma-emitter for SPECT technology (Lagunas-Solar et al., 1981).  $^{205}$Bi and $^{206}$Bi have also medical interest, while they are used to study biokinetics of medically interesting $^{212,213}$Bi alpha-emitter radioisotopes \citep{Milenic}.  Lead and lead-bismuth alloy \citep{Broome} are considered as target material for spallation neutron sources, it is also used for Pb or Pb-Bi, Pb-Mg cooled fast nuclear reactors and lead cooled accelerator driven reactors \citep{IAEAreactor}. The beam handling and target systems (collimator, energy degrader, target backing) are frequently made from lead (brass). At charged particle accelerators, widely used different alloys are also containing lead.
In most of these applications high intensity, low and high energy direct or secondary proton beams activate these technological elements and produce highly active radio-products. Recognizing the importance of knowledge of the production cross-sections on a large energy scale systematic coordinated experimental and theoretical studies were indicated and large scale experiments were done. In spite of this, only a few experimental data sets exist, collected mostly by the experimental groups. In the low energy region practically only one experimental cross-section is available from the Hannover-Cologne group.
In our research and application work in connection with the activation cross-sections on lead we were involved in different projects and applications: 
\begin{itemize}
\item preparation of nuclear database for production of medical radioisotopes in the frame of an  IAEA Coordinated Research Project \citep{Gul,Zaitseva}, using the  $^{nat}$Pb(p,x)$^{201}$Tl reaction
\item Preparation of  proton and deuteron activation cross-section database for FENDL Fusion Evaluated Data Library \citep{IAEAFENDL}
\item Investigating the activation cross-sections of deuteron induced reactions on lead \citep{Ditroi}
\item Preparation of data base of Thin Layer Activation (TLA) technique for wear measurement \citep{IAEANDS} via the $^{nat}$Pb(p,x)$^{206}$Bi  reaction
\item Every day practice of wear measurement of brass samples \citep{Laguzzi}
\end{itemize}
The importance of the cross-section data for nuclear databases was justified in preparation phase in practice by selecting important targets for the database. From other side, in our practice the importance of cross-section data is well known for choosing the appropriate irradiation parameters (irradiation time, energy, intensity, etc.).
Therefore we decided to re-measure the excitation function in the energy range of our cyclotrons ($E_{max}$ = 40 MeV).  According to our experiences, before the experiment we have investigated the earlier experimental results and the theoretical results available in web (TENDL and MENDL2p data-bases, EMPIRE 3.1). We found only few experimental results, out of them results from reliable research groups.

\section{Earlier investigations}
\label{2}
Taking into account the importance of the proton induced activation data on lead the activation cross-sections were measured up to the GeV energies. But due to the used measuring technology (stacked foil irradiation technique, high activities and long cooling time) the cross-sections of the relatively short-lived activation products were not measured. Comparing the available experimental data on other important materials in low energy region the data sets on lead are rather poor. Considering the cross-section and yield data:  
Bell et al. measured the cross-sections of (p,xn) reactions on the isotopes of lead and bismuth up to 50 MeV in 1956 \citep{Bell}. Each target isotope was measured on enriched targets of different compositions and different energy intervals, so a calculation of natural and/or isotopic cross-section for a particular energy was difficult or not possible.
Lagunas-Solar reported differential yield for production of $^{201}$Tl  in 1981 in the 47-6 MeV energy range \citep{Lagunas1981}.  
Production of $^{205,206}$Bi for environmental toxicology and proton activation analysis were investigated in 1981 \citep{Birattari}.
 
Radioactive nuclide yields on thick target for $^{205,206,207}$Bi at 22 MeV from systematical measurement were reported by Dmitriev in 1981 \citep{Dmitriev}.

 Induced radioactivity of component materials by 16 MeV was investigated by Abe et al. 1984 \citep{Abe} .          
The Dubna group investigated the $^{201}$Tl production from natural and enriched lead targets by protons of energy about 65 MeV \citep{Ageev,Zaitseva}.
The Hanover-Cologne group in series of measurements investigated the activation cross-sections at different accelerators up to high energies. In the low energy measurement by  Kuhnhenn et al.   a detailed study was performed for series of activation products  \citep{Kuhnhenn}.	
Milazzo-Colli \citep{Milazzo} investigated (p,$\gamma$) reaction on  deposited Pb samples resulting in such Tl isotopes, which are out of the scope of the present study.

\section{Experimental and data evaluation}
\label{3}
Excitation functions were measured by stacked foil irradiation technique at the VUB CGR-560 cyclotron (Brussels). The stacks were irradiated in a Faraday-cup like target holder, equipped with a long collimator for precise definition of the diameter of the incident beam and with an electrode to suppress the secondary electrons emitted from the target. A stack of 12 identical groups of (Ho-Zr-Pb-Ti-Al) foil layers containing holmium (26.2 $\mu m$), zirconium (98.42 $\mu m$), lead  (15.74 $\mu m$) target foils, Ti (12 $\mu m$) beam monitor foil and aluminum (156.6 $\mu m$) beam energy degrader foil  of natural isotopic composition was irradiated with a proton beam of 37 MeV primary energy and 146 nA intensity for 70 min. The beam current was kept constant during the irradiation. The activity of the irradiated samples was measured directly, without any further chemical separation by a high purity Ge detector. The target samples were counted four times after 1.3 h, 3.5 h, 97 h and 285 h typical cooling times, in order to detect all possible radioisotopes according to their half-life as well as to follow the decay of radioisotopes detected in more than one measurement.
 As naturally occurring lead is composed of four stable isotopes ($^{204}$Pb 1.4\%; $^{206}$Pb 24.1\%; $^{207}$Pb 22.1\%; $^{208}$Pb 52.4\%), so called elemental cross-sections were determined supposing the lead is composed of only one single isotope. In the final result cross-section for natural composition is given. The radionuclides of Bi, Pb and Tl are produced by (p,xn), (p,pxn) and (p,2pxn) reactions respectively and through the decay of their parent nuclei. The decay characteristics, the contributing reactions/processes and the respective reaction Q-values for production of the measured radioisotopes were taken from the NUDAT2 data-base \citep{Kinsey} and from \citep{Q} and are summarized in Table 1. In the same Table the $\gamma$-lines with their intensities are also reported. All measurable lines were evaluated and those having appropriate statistics and no interference have been used in the final calculations.
The initial value of the primary bombarding energy of the incident particles was derived from calibrations made with TOF technique \citep{Sonck}. The energy degradation of the protons through the stack was estimated by calculation using the stopping power data of Andersen and Ziegler \citep{Andersen}. The total charge on target was initially derived from the Faraday cup using a digital integrator. The final beam current and energy scale are based on the analysis of the cross-section data for the $^{nat}$Ti(p,x)$^{48}$V monitor reaction measured simultaneously with the investigated reactions \citep{TF1991}. The recommended cross-section data for the reactions were taken from upgraded database of an IAEA research project \citep{TF2001}. The agreement of the cross-sections of the monitor reaction with the recommended cross-section (measured over a whole energy range) is shown in Fig. 1.
The uncertainty on each cross-section data point was estimated in a standard way \citep{Error} by taking the square root of the sum of the square of all relative individual contributions (except the nonlinear time parameters), supposing equal sensitivities for the different parameters appearing in the formula. The final uncertainties of the cross-sections contain uncertainties of the beam current measurement (7 \%), the number of target nuclei (5 \%), the determination of activities and conversion to absolute number of the produced nuclei (1-15 \%). The absolute values of the cross- sections are estimated to be accurate within 12 \%. 
The uncertainty of the energy scale was estimated by taking into account the energy uncertainty of the primary beam (0.3 MeV), the possible variation in the target thickness and the effect of beam straggling. The energy uncertainty in the last foil was estimated to be 0.75 MeV.

\begin{table*}[t]
\tiny
\caption{Decay characteristic and contributing reactions for the investigated radioisotopes $^{206,205,204,203, 202,201g}$Bi,  $^{203cum,202m,201cum}$Pb and $^{202cum,201cum,200cum,199cum}$Tl }
\centering
\begin{center}
\begin{tabular}{|p{0.7in}|p{0.5in}|p{0.6in}|p{0.5in}|p{0.9in}|p{0.8in}|} \hline 
\textbf{Nuclide} & \textbf{Half-life} & \textbf{E$_{\gamma}$(keV)} & \textbf{I$_{\gamma}$(\%)} & \textbf{Contributing reactions} & \textbf{Q-value\newline (keV)} \\ \hline 
\textbf{${}^{207}$Bi\newline }$\varepsilon $: 100~\%\textbf{} & 31.55 a & ~569.698\newline 1063.656 & ~97.75 \newline 74.5 & ${}^{207}$Pb(p,n)\newline ${}^{208}$Pb(p,2n) & ~-3179.76\newline ~-10547.63 \\ \hline 
\textbf{${}^{206}$Bi\newline }$\varepsilon $: 100~\%\textbf{} & 6.243 d & ~183.977\newline ~343.51\newline 398.00\newline 497.06\newline 516.18\newline 537.45\newline 803.10\newline 881.01\newline 895.12\newline 1098.26 & ~15.8\newline ~23.5~\newline 10.75\newline 15.33\newline 40.8\newline 30.5\newline 99.0~\newline 66.2~\newline 15.67\newline ~13.51 & ${}^{206}$Pb(p,n)\newline ${}^{207}$Pb(p,2n)\newline ${}^{208}$Pb(p,3n)\newline  & -4539.65\newline -11277.43 \newline -18645.3 \\ \hline 
\textbf{${}^{205}$Bi\newline }~$\varepsilon $: 100~\%\textbf{\newline } & 15.31 d & 703.45\newline 987.66\newline ~1764.30 & 31.1\newline 16.1\newline ~32.5 & ${}^{206}$Pb(p,2n)\newline ${}^{207}$Pb(p,3n)\newline ${}^{208}$Pb(p,4n) & -11574.57\newline -18312.35\newline -25680.21 \\ \hline 
\textbf{${}^{204}$Bi\newline }e: 99.75\%\newline b${}^{+}$:0.25 \% & 11.22 h & 374.76\newline ~670.72\newline 899.15\newline 911.74\newline ~911.96\newline ~918.26\newline 983.98 & ~82\newline 11.4\newline 99\newline 13.6~\newline 11.2\newline ~ 10.9\newline 59~ & ${}^{204}$Pb(p,n)\newline ${}^{206}$Pb(p,3n)\newline ${}^{207}$Pb(p,4n)\newline ${}^{208}$Pb(p,5n)\newline  & ~-5246.14\newline -20064.47\newline -26802.25\newline -34170.11 \\ \hline 
\textbf{${}^{203}$Bi\newline }$\varepsilon $: 100~\%\textbf{\newline } & 11.76 h & 816.3\newline 820.2\newline 825.2\newline 847.2\newline 896.9\newline 1033.7\newline 1679.6 & 4.1\newline 30.0\newline ~14.8\newline 8.6\newline 13.2\newline 8.9 \newline 8.9 & ${}^{204}$Pb(p,2n)\newline ${}^{206}$Pb(p,4n)\newline ${}^{207}$Pb(p,5n)\newline ${}^{208}$Pb(p,6n)\newline  & -12438.8\newline ~-27257.1\newline -33994.9\newline -41362.8 \\ \hline 
\textbf{${}^{202}$Bi\newline }$\varepsilon $: 100~\%~\textbf{\newline } & 1.71 h & 422.13\newline 657.49\newline 960.67 & 83.7\newline 60.6\newline ~99.283 & ${}^{204}$Pb(p,3n)\newline ${}^{206}$Pb(p,5n)\newline ${}^{207}$Pb(p,6n)\newline ${}^{208}$Pb(p,7n) & -21293.6\newline -36111.9\newline -42849.7\newline -50217.5 \\ \hline 
\textbf{${}^{201m}$Bi\newline }~$\alpha $~~0.3 \%\newline $\varepsilon $: 91.1~\%~\textbf{\newline }IT: 8.6 \%\newline 846.34\textit{2 keV}\textbf{\newline } & 59.1 min & 846.4 & 5 & ${}^{204}$Pb(p,4n)\newline ${}^{206}$Pb(p,6n)\newline ${}^{207}$Pb(p,7n)\newline ${}^{208}$Pb(p,8n) & ~-28691.0~\newline ~-43509.3\newline ~-50247.1\newline ~-57615.0 \\ \hline 
\textbf{${}^{201g}$Bi\newline }$\varepsilon $: 100~\%\textbf{} & 103 min & 629.1~\newline 786.4\newline ~936.2\newline 1014.1 & ~26.0\newline 10.3\newline 12.2\newline 11.6 & ${}^{204}$Pb(p,4n)\newline ${}^{206}$Pb(p,6n)\newline ${}^{207}$Pb(p,7n)\newline ${}^{20}$Pb(p,8n) & ~-28691.0~\newline ~-43509.3\newline ~-50247.1\newline ~-57615.0 \\ \hline 
\textbf{${}^{203}$Pb\newline }$\varepsilon $: 100~\%\textbf{} & 51.92 h & 279.1952 & 80.9 & ${}^{204}$Pb(p,pn)\newline ${}^{206}$Pb(p,p3n)\newline ${}^{207}$Pb(p,p4n)\newline ${}^{208}$Pb(p,p5n)\newline ${}^{203}$Bi decay & ~~-8394.68~\newline -23213.01\newline -29950.78\newline -37318.65 \\ \hline 
\textbf{${}^{202m}$Pb\newline }$\varepsilon $: 9.5~\%\textbf{\newline }IT: 90.5~\%\newline 2169.83 keV\textbf{} & 3.54 h & 422.12\newline 657.49\newline 786.99\newline 960.70 & 84\newline 31.7\newline 49\newline 89.9~ & ${}^{204}$Pb(p,p2n)\newline ${}^{206}$Pb(p,p4n)\newline ${}^{207}$Pb(p,p5n)\newline ${}^{208}$Pb(p,p6n) & -15311.78\newline -30130.11\newline -36867.88\newline -44235.76 \\ \hline 
\textbf{${}^{201}$Pb\newline }$\varepsilon $~:100 \%\textbf{} & ~9.33 h & 331.15~\newline 907.67\newline 945.96 & 77~\newline 6.~7.21\newline ~ & ${}^{204}$Pb(p,p3n)\newline ${}^{206}$Pb(p,p5n)\newline ${}^{207}$Pb(p,p6n)\newline ${}^{208}$Pb(p,p7n)\newline ${}^{20}$${}^{1}$Bi decay & ~-24063.9\newline -38882.2\newline -45620.0\newline -52987.9 \\ \hline 
\textbf{${}^{202}$Tl\newline }$\varepsilon $: 100~\%~\textbf{} & ~12.31 d & 39.510~ & 91.5 & ${}^{204}$Pb(p,2pn)\newline ${}^{206}$Pb(p,2p3n)\newline ${}^{207}$Pb(p,2p4n)\newline ${}^{208}$Pb(p,2p5n)\newline ${}^{202}$Pb decay & -14483.2\newline -29301.6\newline ~-36039.4\newline ~-43407.2 \\ \hline 
\textbf{${}^{201}$Tl\newline }$\varepsilon $: 100~\%~\textbf{} & 3.0421 d & 135.34~\newline ~167.43 & 2.565\newline ~10.00 & ${}^{204}$Pb(p,2p2n)\newline ${}^{206}$Pb(p,2p4n)\newline ${}^{207}$Pb(p,2p5n)\newline ${}^{208}$Pb(p,2p6n)\newline ${}^{201}$Pb decay & ~-21362.0\newline -36180.4\newline -42918.1\newline -50286.0 \\ \hline 
\textbf{${}^{200}$Tl\newline }$\varepsilon $: 100~\%\textbf{} & 26.1 h~ & ~367.942\newline 579.300\newline 828.27\newline ~1205.75 & ~87\newline 13.7\newline 10.8~\newline 30~ & ${}^{204}$Pb(p,2p3n)\newline ${}^{206}$Pb(p,2p5n)\newline ${}^{207}$Pb(p,2p6n)\newline ${}^{208}$Pb(p,2p7n) & -29564.74\newline -44383.07\newline ~-51120.85\newline -58488.72 \\ \hline 
\textbf{${}^{199}$Tl\newline }$\varepsilon $: 100~\%\textbf{} & 7.42 h\newline  & 158.359\newline 208.20\newline 247.26\newline 455.46~ & 5.0\newline 12.3\newline ~9.3\newline ~12.4 & ${}^{204}$Pb(p,2p4n)\newline ${}^{206}$Pb(p,2p6n)\newline ${}^{207}$Pb(p,2p7n)\newline ${}^{208}$Pb(p,2p8n)\newline ${}^{199}$Pb decay & ~-36624.2\newline -51442.5\newline -58180.3\newline ~-65548.2\newline  \\ \hline 
\end{tabular}

\end{center}
\begin{flushleft}
\footnotesize{\noindent When complex particles are emitted instead of individual protons and neutrons the Q-values have to be modified by the respective binding energies of the compound particles: np-d, +2.2 MeV; 2np-t, +8.48 MeV; 2p2n-a, 28.30 MeV.

\noindent When complex particles are emitted instead of individual protons and neutrons the Q-values have to be decreased by the respective binding energies of the compound particles: np-d, +2.2 MeV; 2np-t, +8.48 MeV; n2p-${}^{3}$He, +7.72 MeV; 2n2p-$\alpha$, +28.30 MeV
\noindent *Decrease Q-values for isomeric states with level energy of the isomer}
\end{flushleft}

\end{table*}

\begin{figure}
\includegraphics[width=0.5\textwidth]{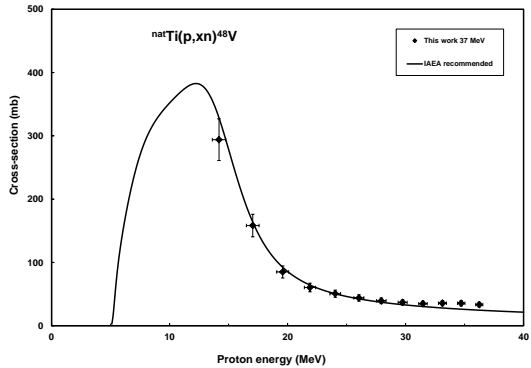}
\caption{Measured cross-section of the $^{nat}$Ti(p,xn)$^{48}$V monitor reaction, comparison with the recommended excitation function}
\label{fig:1}       
\end{figure}

\section{Results}
\label{4}

\subsection{Cross-sections}
\label{4.1}
The excitation functions for production of radioisotopes of Bi, Pb and Tl are shown in Figs. 2-14 in comparison with the earlier experimental data and with the theoretical results in TENDL-2012  \citep{KoningTENDL} library based on TALYS model code \citep{KoningTALYS} as well as with the results of the EMPIRE 3.1 \citep{Herman2007,Herman2012} and ALICE-IPPE (MENDL2p) \citep{IAEAMENDL} calculations. TENDL-library already contains the parameter-adjusted results of the TALYS results, while EMPIRE and ALICE-IPPE run in blind mode without parameter adjustment. For applications, the new experimental cross-sections for radionuclides produced are tabulated in Tables 2 and 3.
The "m+" notes the activation cross-section of the ground-state after the "complete" decay of the significantly shorter-lived isomeric state decaying partly or completely to that ground state and the "cum" notes the activation cross-section of the final product after the "complete" decay of the simultaneously produced significantly shorter-lived parent nuclei decaying partly or completely to the investigated final product.
The available results of earlier investigations published in the literature are discussed together with results of theoretical calculations.

\subsubsection{$^{nat}$Pb(p,xn)$^{206}$Bi reaction}
\label{4.1.1}
The radioisotope $^{206}$Bi ($T_{1/2}$ = 6.243 d) is produced by the (p,n), (p,2n) and (p,3n) reactions from $^{206}$Pb, $^{207}$Pb and $^{208}$Pb lead isotopes respectively. Its relatively short half-life makes quick measurements possible within a couple of days. Our new experimental results are presented in Fig. 2 together with the earlier experimental values from the literature and the results of the theoretical model calculations. Our new data are in good agreement with the results of Kuhnhenn \citep{Kuhnhenn}, Bell \citep{Bell} above 30 MeV and Lagunas-Solar \citep{Lagunas1987} under 22 and above 34 MeV. The results of Lagunas-Solar are too low around the maximum of the excitation function curve, while the data of Bell are too low below 30 MeV. TENDL-2012 and MENDL give good approximation except for the maximum value, which was a little overestimated by both codes. The EMPIRE 3.1 results strongly underestimate the experimental values, which can be explained by the fact that the EMPIRE input library (RIPL3) was not complete for the high-Z target elements, so data-extrapolations should have been made to avoid EMPIRE to fail because of fatal errors.	

\begin{figure}
\includegraphics[width=0.5\textwidth]{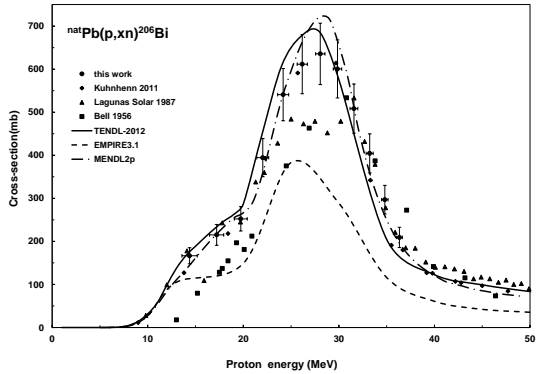}
\caption{Excitation function of the $^{nat}$Pb(p,xn)$^{206}$Bi reaction}
\label{fig:2}       
\end{figure}


\begin{figure}
\includegraphics[width=0.5\textwidth]{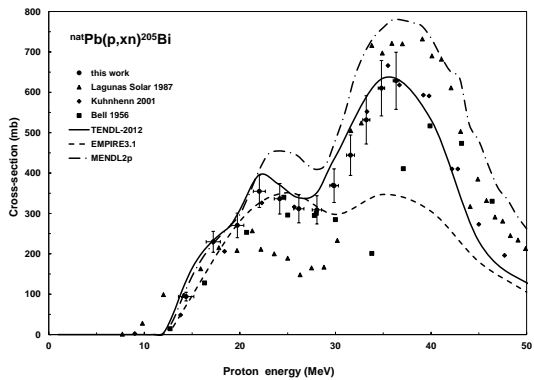}
\caption{Excitation function of the $^{nat}$Pb(p,xn)$^{205}$Bi reaction}
\label{fig:3}       
\end{figure}

\subsubsection{$^{nat}$Pb(p,xn)$^{205}$Bi reaction}
\label{4.1.2}
The radioisotope $^{205}$Bi ($T_{1/2}$ = 15.31 d) is produced in a similar way from the same Pb isotopes as the $^{206}$Pb above, but with one more neutron emission respectively. Its half-life is long enough to follow its decay through several weeks. Our new experimental data are presented in Fig. 3 together with the literature values and with the results of theoretical model calculations. Now there is acceptable agreement with the earlier data of Kuhnhenn \citep{Kuhnhenn}. The data of Lagunas-Solar \citep{Lagunas1987} strongly decline both from ours and from the data of Kuhnhenn especially around the both local maxima of the curve. The results of Bell \citep{Bell} agree acceptably below 25 MeV and above 40 MeV (with other previous measurements), but too low around the second maximum. The best theoretical approximation is given by the TEND-2012, MENDL follows its shape by a little overestimation, so does EMPIRE 3.1 up to 28 MeV, but it strongly underestimates above this energy.

\subsubsection{$^{nat}$Pb(p,xn)$^{204}$Bi reaction }
\label{4.1.3}
The radioisotope $^{204}$Bi ($T_{1/2}$ = 11.22 h) is produced in a similar way from the same Pb isotopes as the above two Bi isotopes with further neutron emissions. Because of its short half-life quick measurements are required. The results of our new measurements are presented in Fig. 4, compared with the data from the literature and the results of the theoretical model calculations. Our results are in acceptable agreement with the values of Kuhnhenn \citep{Kuhnhenn}  again, the data of Bell \citep{Bell}  are significantly lower than ours above 25 MeV. TENDL-2012 and MENDL give acceptable approximation of the experimental values including both our and the literature data. EMPIRE 3.1 strongly underestimates again. 	

\begin{figure}
\includegraphics[width=0.5\textwidth]{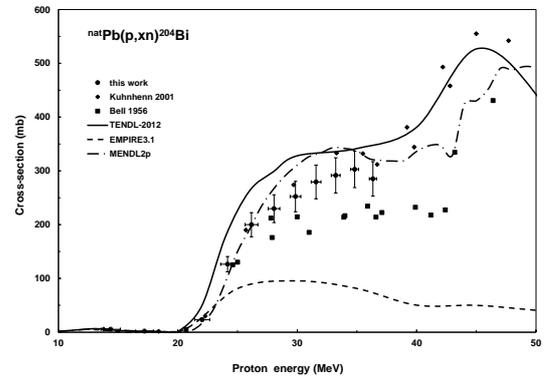}
\caption{Excitation function of the $^{nat}$Pb(p,xn)$^{204}$Bi reaction}
\label{fig:4}       
\end{figure}

\subsubsection{$^{nat}$Pb(p,xn)$^{203}$Bi reaction}
\label{4.1.4}
The radioisotope $^{203}$Bi ($T_{1/2}$ = 11.76 h) is produced from every stable Pb isotope through multiple neutron emission. Its half-life is similarly short as by the $^{204}$Bi above so it also requires quick measurements shortly after the irradiation. Our new data, the literature values and the results of the theoretical model calculations are presented in Fig 5. Now the data of Kuhnhenn \citep{Kuhnhenn}  are much larger than ours, especially above 20 MeV. The TENDL-2012 and the MENDL results follow our experimental values quite well, while EMPIRE 3.1 underestimates again. 

\begin{figure}
\includegraphics[width=0.5\textwidth]{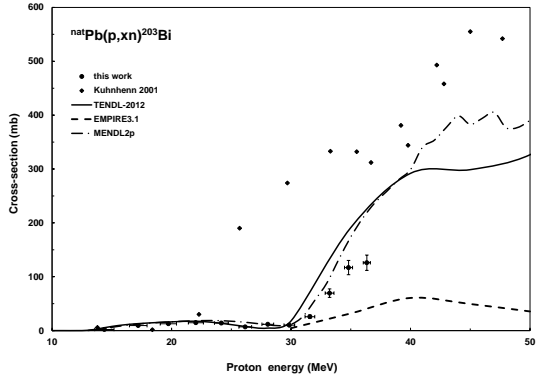}
\caption{Excitation function of the $^{nat}$Pb(p,xn)$^{203}$Bi reaction}
\label{fig:5}       
\end{figure}

\subsubsection{$^{nat}$Pb(p,xn)$^{202}$Bi reaction}
\label{4.1.5}
The radioisotope $^{202}$Bi ($T_{1/2}$ = 1.71 h) is also produced from every stable Pb isotope through multiple neutron emission. Its half-life is very short, so it must be measured just after the irradiation. The results of this work and the results of the theoretical model calculations are presented in Fig 6. TENDL-2012 and MENDL give acceptable approximation in the energy range up to 37 MeV. EMPIRE 3.1 fails again. No earlier measurement was found for this isotope in the investigated energy interval, so our present results are new. Data found from Bell \citep{Bell}  seem to be a good extrapolation of our new results toward higher energies.

\begin{figure}
\includegraphics[width=0.5\textwidth]{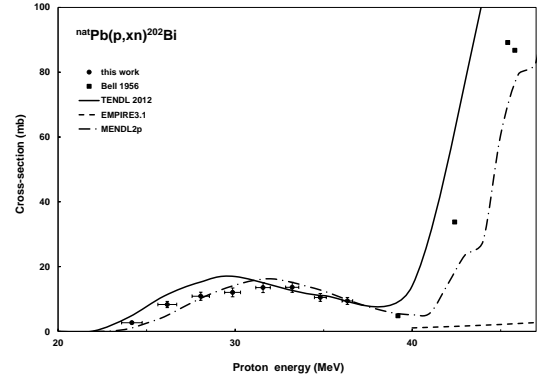}
\caption{Excitation function of the $^{nat}$Pb(p,xn)$^{202}$Bi reaction}
\label{fig:6}       
\end{figure}

\subsubsection{$^{nat}$Pb(p,xn)$^{201g}$Bi reaction }
\label{4.1.6}
The radioisotope $^{201}$Bi ($T_{1/2}$ = 103 min, Tm1/2 = 59.1 min) has a short-lived isomeric state and a bit longer-lived ground state.  We could measure only two data points, which are under the TENDL-2012 and MENDL curve and above the EMPIRE 3.1 curve (see Fig. 7). The half-lives of the isomeric and ground states are similar, so the results cannot be called "m+" as the cross-section after the complete decay of the isomeric state.

\begin{figure}
\includegraphics[width=0.5\textwidth]{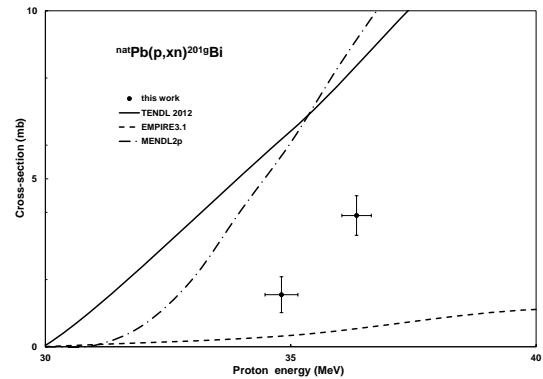}
\caption{Excitation function of the $^{nat}$Pb(p,xn)$^{201g}$Bi reaction}
\label{fig:7}       
\end{figure}

\subsubsection{$^{nat}$Pb(p,X)$^{203}$Pb reaction}
\label{4.1.7}
The radioisotope $^{203}$Pb ($T_{1/2}$ = 51.92 h) is produced from the Pb stable isotopes through one proton and multiple neutron emission, including complex particle emissions (d, t)  after the proton bombardment, as well as from the decay of the parallel produced $^{203}$Bi mother isotope. The measurements were performed by keeping appropriate cooling time (the difference between the mother's and the daughter's half-life allowed it) in order to measure cumulative cross-section after the complete decay of the mother isotope. The results are presented in Fig. 8 together with the earlier experimental data and the results of theoretical model calculations. The agreement with the cumulative results of Kuhnhenn \citep{Kuhnhenn}  is acceptable, the TENDL-2012 approximation (cumulative) is also good, while both EMPIRE 3.1 and MENDL fail in this case because of strong underestimation. The previous results of Bell \citep{Bell}	 are completely different from all other data.

\begin{figure}
\includegraphics[width=0.5\textwidth]{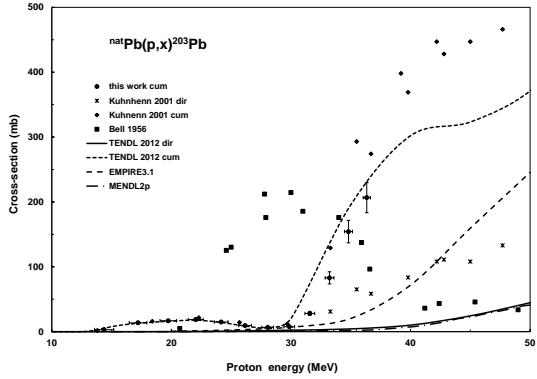}
\caption{Excitation function of the $^{nat}$Pb(p,xn)$^{203}$Pb reaction}
\label{fig:8}       
\end{figure}

\subsubsection{$^{nat}$Pb(p,X)$^{202m}$Pb reaction}
\label{4.1.8}
The radioisotope $^{202}$Pb has a very long-lived ground state and a 3.54 h isomeric state (measured). No earlier measurements were found in the literature. The results are presented in Fig. 9, together with the results of the theoretical model calculations. The data are scattered because of the low counting statistics, where the peak area was comparable with the background fluctuations. In this case, all theoretical model calculations show strong underestimations. 

\begin{figure}
\includegraphics[width=0.5\textwidth]{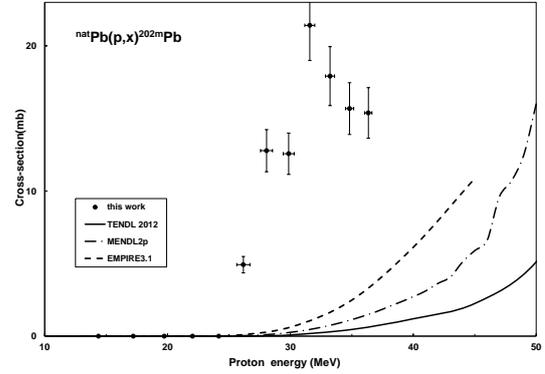}
\caption{Excitation function of the $^{nat}$Pb(p,xn)$^{202m}$Pb reaction}
\label{fig:9}       
\end{figure}

\subsubsection{$^{nat}$Pb(p,X)$^{201}$Pb reaction}
\label{4.1.9}
The radioisotope $^{201}$Pb ($T_{1/2}$ = 9.33 h) is produced from the stable Pb isotopes by one proton and multiple neutron emission, including complex particle emissions (d, t), as well as from the decay of the parallel produced $^{201}$Bi mother isotope. Because of the large difference between the mother's and the daughter's half-lives, cumulative cross-section could be measured after appropriate cooling time. Now only one data point of Kuhnhenn \citep{Kuhnhenn}  overlaps in energy with our measurement and it is in agreement with our new data, while above 36 MeV (where we have no measurement points) the Kuhnhenn data go up to a 5 times larger range. TENDL-2012 gives good approximation, while MENDL and EMPIRE 3.1 strongly underestimate the experimental values. The results are shown in Fig 10.	

\begin{figure}
\includegraphics[width=0.5\textwidth]{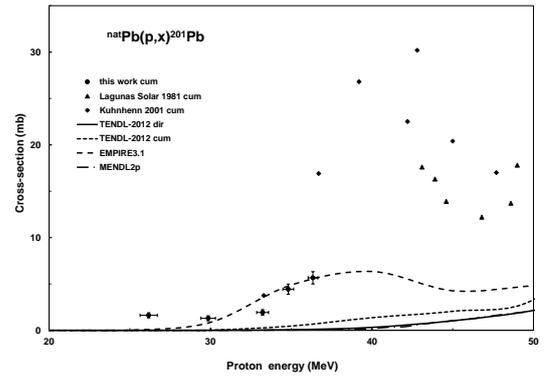}
\caption{Excitation function of the $^{nat}$Pb(p,xn)$^{201}$Pb reaction}
\label{fig:10}       
\end{figure}

\subsubsection{$^{nat}$Pb(p,X)$^{202}$Tl reaction}
\label{4.1.10}
The radioisotope $^{202}$Tl ($T_{1/2}$ = 12.31 d) is produced from the stable Pb isotopes by multiple particle emission, including complex particles (p,n,d,t,$^3$He,$\alpha$), as well as from the decay of the much shorter-lived 202Pb mother isotope.  We measured after the decay of the mother isotope (cumulative cross-section) only 3 data points with relatively large errors, but they are in acceptable agreement with the earlier results of Kuhnhenn \citep{Kuhnhenn} in the selected energy region. Also TENDL-2012 and MENDL give acceptable approximations, but EMPIRE 3.1 strongly overestimates the experimental values (see Fig. 11). 

\begin{figure}
\includegraphics[width=0.5\textwidth]{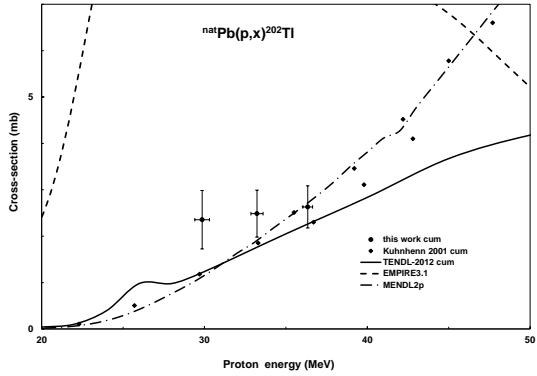}
\caption{Excitation function of the $^{nat}$Pb(p,xn)$^{202}$Tl reaction}
\label{fig:11}       
\end{figure}

\subsubsection{$^{nat}$Pb(p,X)$^{201}$Tl reaction}
\label{4.1.11}
The radioisotope $^{201}$Tl ($T_{1/2}$ = 3.0421 d) is produced in a similar way as the $^{202}$Tl above. Our measured data points are in acceptable agreement with those of Kuhnhenn \citep{Kuhnhenn}. In this case none of the theoretical model codes could provide an acceptable approximation. The results are presented in Fig. 12.

\begin{figure}
\includegraphics[width=0.5\textwidth]{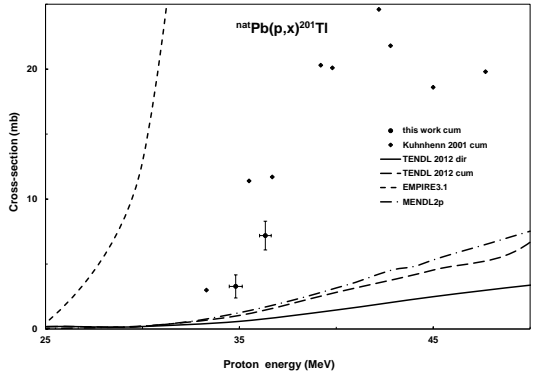}
\caption{Excitation function of the $^{nat}$Pb(p,xn)$^{201}$Tl reaction}
\label{fig:12}       
\end{figure}

\subsubsection{$^{nat}$Pb(p,X)$^{200}$Tl reaction}
\label{4.1.12}
The radioisotope $^{200}$Tl ($T_{1/2}$ = 26.1 h) is produced in a similar way as the $^{202,201}$Tl above. The only difference that the half-lives of the mother and daughter isotopes are comparable, so the cross-sections must be corrected for the mothers contribution. The results are presented in Fig. 13, together with the earlier measurements and the results of theoretical model calculations. Unfortunately there is no overlapping energy region with earlier experimental data, but it also means that our data are new. Surprisingly the best (but still not acceptable) approximation is given by the EMPIRE 3.1, while the results of TENDL-2012 and MENDL cannot be evaluated. 

\begin{figure}
\includegraphics[width=0.5\textwidth]{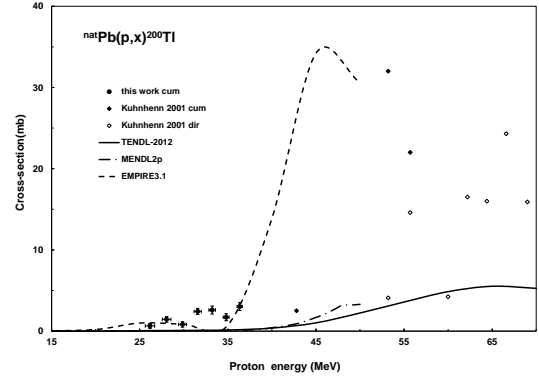}
\caption{Excitation function of the $^{nat}$Pb(p,xn)$^{200}$Tl reaction}
\label{fig:13}       
\end{figure}

\subsubsection{$^{nat}$Pb(p,X)$^{199}$Tl reaction}
\label{4.1.13}
The radioisotope $^{199}$Tl ($T_{1/2}$ = 7.42 h) is produced in a similar way as the $^{202,201,200}$Tl above. Now the half-life of the mother $^{199}$Pb (if produced) is really shorter, than that of the $^{199}$Tl, so the measured cross-section is cumulative, when it is measured after appropriate cooling time (8-10 h). The results are presented in Fig. 14, together with the results of the theoretical model calculations. The approximation given by EMPIRE 3.1 is the best again, while TENDL-2012 and MENDL completely fail because of large underestimation.

\begin{figure}
\includegraphics[width=0.5\textwidth]{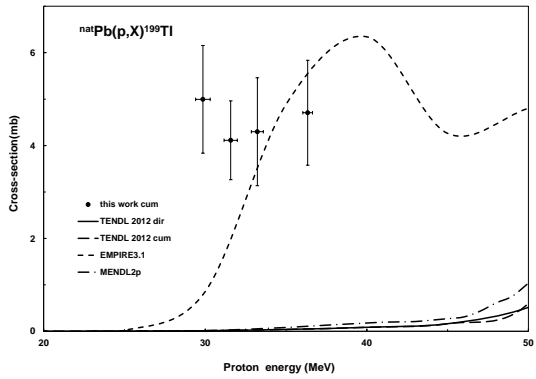}
\caption{Excitation function of the $^{nat}$Pb(p,xn)$^{199}$Tl reaction}
\label{fig:14}       
\end{figure}

\begin{table*}[t]
\tiny
\caption{Measured cross-sections of the ${}^{nat}$Pb(p,xn)${}^{201}$${}^{g,202,203,204,205,206}$Bi nuclear reactions}
\centering
\begin{center}
\begin{tabular}{|p{0.2in}|p{0.2in}|p{0.4in}|p{0.3in}|p{0.3in}|p{0.3in}|p{0.3in}|p{0.3in}|p{0.3in}|p{0.3in}|p{0.3in}|p{0.3in}|p{0.3in}|p{0.3in}|} \hline 
\multicolumn{2}{|p{0.4in}|}{ E $\pm\Delta$E (MeV)} & \multicolumn{12}{|p{4.8in}|}{ Cross-section $\sigma\pm\Delta\sigma$ (mb)} \\ \hline 
\multicolumn{2}{|p{0.6in}|}{} & \multicolumn{2}{|p{0.6in}|}{${}^{206}$Bi} & \multicolumn{2}{|p{0.6in}|}{${}^{205}$Bi} & \multicolumn{2}{|p{0.6in}|}{${}^{204}$Bi} & \multicolumn{2}{|p{0.6in}|}{${}^{203}$Bi${}^{  }$} & \multicolumn{2}{|p{0.6in}|}{${}^{202}$Bi} & \multicolumn{2}{|p{0.6in}|}{${}^{201g}$Bi} \\ \hline 
14.4 & 0.8 & 166.6 & 18.7 & 94.1 & 10.8 & 5.4 & 0.6 & 2.1 & 0.4 & ~ &  & ~ &  \\ \hline 
17.2 & 0.7 & 215.1 & 24.2 & 229.7 & 26.0 & 2.0 & 0.3 & 9.4 & 1.1 & ~ &  & ~ &  \\ \hline 
19.7 & 0.7 & 252.3 & 28.3 & 270.4 & 30.5 & ~ &  & 12.7 & 1.5 & ~ &  & ~ &  \\ \hline 
22.0 & 0.6 & 394.5 & 44.3 & 354.8 & 40.0 & 23.4 & 2.6 & 14.6 & 1.7 & ~ &  & ~ &  \\ \hline 
24.2 & 0.6 & 540.8 & 60.7 & 336.4 & 37.8 & 126.5 & 14.2 & 14.0 & 1.7 & 2.7 & 0.3 & ~ &  \\ \hline 
26.2 & 0.5 & 611.5 & 68.7 & 311.6 & 35.1 & 200.0 & 22.5 & 6.9 & 0.8 & 8.3 & 1.0 & ~ &  \\ \hline 
28.1 & 0.5 & 635.6 & 71.4 & 308.9 & 35.0 & 229.7 & 25.8 & 11.7 & 1.5 & 10.9 & 1.2 & ~ &  \\ \hline 
29.9 & 0.4 & 600.5 & 67.4 & 368.7 & 41.6 & 252.3 & 28.3 & 10.0 & 1.3 & 12.0 & 1.4 & ~ &  \\ \hline 
31.6 & 0.4 & 508.5 & 57.1 & 444.3 & 50.0 & 279.4 & 31.4 & 25.8 & 3.0 & 13.5 & 1.5 & ~ &  \\ \hline 
33.2 & 0.4 & 404.4 & 45.4 & 531.7 & 59.8 & 291.6 & 32.8 & 69.5 & 7.9 & 13.6 & 1.5 & ~ &  \\ \hline 
34.8 & 0.3 & 296.9 & 33.3 & 610.2 & 68.7 & 302.9 & 34.0 & 117.0 & 13.2 & 10.5 & 1.2 & 1.5 & 0.5 \\ \hline 
36.3 & 0.3 & 209.2 & 23.5 & 628.6 & 70.7 & 285.2 & 32.0 & 126.0 & 14.2 & 9.4 & 1.1 & 3.9 & 0.6 \\ \hline 
\end{tabular}

\end{center}
\end{table*}

\begin{table*}[t]
\tiny
\caption{Measured cross-sections of the ${}^{nat}$Pb(p,X)${}^{201,202m,203}$Pb and the ${}^{nat}$Pb(p,X)${}^{199,200,201,202}$Tl nuclear reactions}
\centering
\begin{center}
\begin{tabular}{|p{0.2in}|p{0.2in}|p{0.3in}|p{0.3in}|p{0.3in}|p{0.3in}|p{0.3in}|p{0.3in}|p{0.3in}|p{0.3in}|p{0.3in}|p{0.3in}|p{0.3in}|p{0.3in}|p{0.3in}|p{0.3in}|} \hline 
\multicolumn{2}{|p{0.4in}|}{ E $\pm\Delta$E (MeV)} & \multicolumn{14}{|p{4.2in}|}{ Cross-section $\sigma\pm\Delta\sigma$ (mb)} \\ \hline 
\multicolumn{2}{|p{0.6in}|}{} & \multicolumn{2}{|p{0.6in}|}{$^{203}$Pb} & \multicolumn{2}{|p{0.6in}|}{$^{202m}$Pb} & \multicolumn{2}{|p{0.6in}|}{$^{201}$Pb} & \multicolumn{2}{|p{0.6in}|}{$^{202}$Tl} & \multicolumn{2}{|p{0.6in}|}{$^{201}$Tl} & \multicolumn{2}{|p{0.6in}|}{$^{200}$Tl} & \multicolumn{2}{|p{0.6in}|}{$^{199}$Tl} \\ \hline 
14.4 & 0.8 & 3.1 & 0.4 & ~ &  & ~ &  & ~ &  & ~ &  & ~ &  & ~ &  \\ \hline 
17.2 & 0.7 & 13.7 & 1.5 & ~ &  & ~ &  & ~ &  & ~ &  & ~ &  & ~ &  \\ \hline 
19.7 & 0.7 & 16.8 & 1.9 & ~ &  & ~ &  & ~ &  & ~ &  & ~ &  & ~ &  \\ \hline 
22.0 & 0.6 & 18.7 & 2.1 & ~ &  & ~ &  & ~ &  & ~ &  & ~ &  & ~ &  \\ \hline 
24.2 & 0.6 & 14.9 & 1.7 & ~ &  & ~ &  & ~ &  & ~ &  & ~ &  & ~ &  \\ \hline 
26.2 & 0.5 & 9.4 & 1.1 & 4.9 & 0.6 & 1.6 & 0.3 & ~ &  & ~ &  & 0.6 & 0.3 & ~ &  \\ \hline 
28.1 & 0.5 & 6.6 & 0.8 & 12.8 & 1.5 & ~ &  & ~ &  & ~ &  & 1.4 & 0.3 & ~ &  \\ \hline 
29.9 & 0.4 & 7.6 & 0.9 & 12.6 & 1.4 & 1.3 & 0.3 & 2.4 & 0.6 & ~ &  & 0.8 & 0.3 & 5.0 & 1.2 \\ \hline 
31.6 & 0.4 & 28.1 & 3.2 & 21.4 & 2.4 & ~ &  & ~ &  & ~ &  & 2.4 & 0.4 & 4.1 & 0.8 \\ \hline 
33.2 & 0.4 & 83.0 & 9.3 & 17.9 & 2.0 & 1.9 & 0.3 & 2.5 & 0.5 & ~ &  & 2.6 & 0.5 & 4.3 & 1.2 \\ \hline 
34.8 & 0.3 & 154.3 & 17.3 & 15.7 & 1.8 & 4.4 & 0.6 & ~ &  & 3.3 & 0.9 & 1.7 & 0.4 & ~ &  \\ \hline 
36.3 & 0.3 & 206.6 & 23.2 & 15.4 & 1.7 & 5.7 & 0.7 & 2.6 & 0.5 & 7.2 & 1.1 & 3.0 & 0.5 & 4.7 & 1.1 \\ \hline 
\end{tabular}

\end{center}
\end{table*}

\subsection{Integral yields}
\label{5.}
The integral yields (thick target yields) calculated from the measured cross-sections are shown in Figs. 15, 16 and 17 in comparison with the literature data (if available). The yields represent so called physical yields for instantaneous irradiation  \citep{Bonardi}. Literature data found: $^{205,206}$Bi at 22 MeV \citep{Dmitriev} , $^{205}$Bi (7.7-67 MeV) \citep{Lagunas1987}, $^{205,206}$Bi (8-44 MeV thin target yield) \citep{Birattari} , $^{206}$Bi at 16 MeV \citep{Abe}. In Fig. 15 the results only for $^{205,206}$Bi are presented, as only for these two radioisotopes was the comparison with earlier measurements possible. As it is seen in Fig. 15, the earlier data of Dmitriev are larger ($^{205}$Bi) or lower ($^{206}$Bi) than ours by about 30\%. The earlier measured yield curves of Lagunas-Solar are in acceptable agreement with our new results above 17 MeV. The results of Birratari et al. \citep{Birattari} are in good agreement with our experimental data up to 20 MeV, but they are somewhat lower above this energy. In Figs. 16 and 17 yield calculations for the rest of the measured isotopes are presented, for which no earlier results were found in the literature. 

\begin{figure}
\includegraphics[width=0.5\textwidth]{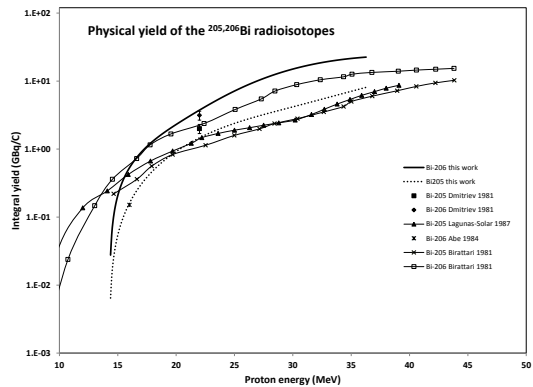}
\caption{Integral yields as a function of proton energy for the production $^{205,206}$Bi radioisotopes compared with the literature}
\label{fig:15}       
\end{figure}

\begin{figure}
\includegraphics[width=0.5\textwidth]{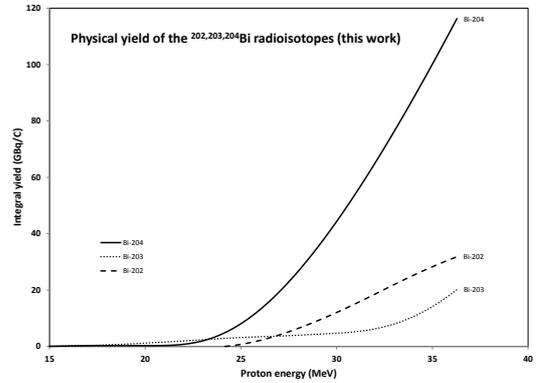}
\caption{Integral yields as a function of proton energy for the production $^{202,203,204}$Bi radioisotopes }
\label{fig:16}       
\end{figure}

\begin{figure}
\includegraphics[width=0.5\textwidth]{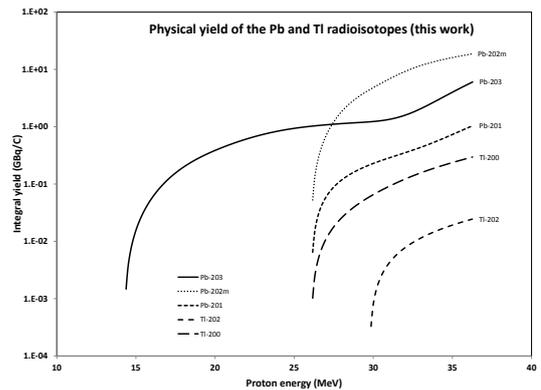}
\caption{Integral yields as a function of proton energy for the production Pb and Tl radioisotopes}
\label{fig:17}       
\end{figure}

\section{Thin Layer Activation (TLA)}
\label{6.}
Wear, corrosion and erosion measurements by using TLA require radioisotopes with appropriate half-lives, intense ?-line and easy production possibility (large cross-section). Among the investigated radionuclides, these conditions are best fulfilled by the $^{205}$Bi and $^{206}$Bi. The cross-sections of both isotopes are similar around 20 MeV, the half-life of the $^{205}$Bi is longer, while $^{206}$Bi has more intense $\gamma$-lines. The IAEA TLA database \citep{IAEANDS} already contained recommended excitation function for the $^{206}$Bi producing reaction. A comparison showed that it is in good agreement with our new measurement. The new $^{205}$Bi results were added to the TLA database and the calculated results are presented in Fig. 18 for special cases. With these limitations, both isotopes are suitable for quick wear measurements by using thin layer activation method. In Fig. 18 we calculated some special cases to demonstrate the wear measurement capability by using these isotopes: Case 1-2 are for $^{205}$Bi, Case 3-4 for $^{206}$Bi optimal, i.e. produce homogenous activity distribution near to the surface.  
Case 1: perpendicular irradiation, Ep = 24.3 MeV, $I_p = 2 \mu A$, T$_{Irradiation}$ = 1 h, T$_{cooling}$ =3 days
Case 2: irradiation angle = $15^o$, Ep = 24.3 MeV, $I_p = 2 \mu A$, T$_{Irradiation}$ = 1 h, T$_{cooling}$ =3 days
Case 3: perpendicular irradiation, Ep = 28.8 MeV, $I_p = 2 \mu A$, T$_{Irradiation}$ = 1 h, T$_{cooling}$ =3 days
Case 4: irradiation angle = $15^o$, Ep = 28.8 MeV, $I_p = 2 \mu A$, T$_{Irradiation}$ = 1 h, T$_{cooling}$ =3 days
From Fig. 18 it is also seen that by angular irradiation ($15^o$) much more dense surface activity distribution can be reached. Important parameters from the point of view of wear measurement is the homogeneous surface activity and the range of homogeneity (a depth range within the activity change is less than 1\%), which are: 21 $kBq/\mu m$ and 192 $\mu m$; 85 $kBq/\mu m$ and 49.8 $\mu m$; 80 $kBq/\mu m$ and 159.6 $\mu m$; 310 $kBq/\mu m$ and 41.3 $\mu m$ for the Cases 1-4 respectively. One can choose from the different irradiation setups according to the requirements (penetration depth, detecting efficiency, expected wear rate, etc.) of the actual measurement. 

\begin{figure}
\includegraphics[width=0.5\textwidth]{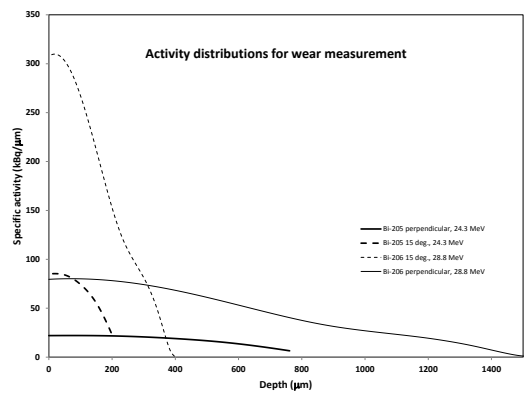}
\caption{Activity distributions for $^{205}$Bi (24.3 MeV) and $^{206}$Bi(28.8 MeV) both for perpendicular and for angular ($15^o$) irradiations: beam current 2 $\mu A$; irradiation time 1 hour; cooling time 3 days }
\label{fig:18}       
\end{figure}

\section{Summary}
\label{7.}
Excitation functions of proton induced nuclear reactions were investigated up to 37 MeV on lead, mainly for application purposes. Independent or cumulative cross-sections for the formation of the radionuclides $^{206,205,204,203,202,201g}$Bi, $^{203cum,202m,201cum}$Pb and $^{202cum,201cum,200cum,199cum}$Tl were measured, out of them the $^{202,201g}$Bi, $^{202m}$Pb and $^{199}$Tl for the first time. 
The new experimental results were compared with the earlier experimental data, with the theoretical result in TENDL-2012, with the results of EMPIRE 3.1 calculations, with the ALICE-IPPE results in the MENDL library and with the recommended data in the IAEA TLA database. Good agreement was found with the experimental data of Kuhnhenn et al. except for $^{203}$Bi. The theoretical predictions in TENDL-2012 and MENDL describe acceptably well the experimental data in most cases, while EMPIRE 3.1 fails except for $^{199,200}$Tl. The new data are in good agreement with the recommended cross-sections for production of $^{206}$Bi in the TLA database.

\section{Acknowledgements}
\label{}
This work was done and supported in the frame of HAS-FWO (Vlaanderen) (Hungary–Belgium) research project.

This work was done in the frame MTA-FWO research project and ATOMKI-CYRIC collaboration. The authors acknowledge the support of research projects and of their respective institutions in providing the materials and the facilities for this work. 
 



\clearpage
\bibliographystyle{elsarticle-harv}
\bibliography{Pbp}







\end{document}